\documentclass[fleqn,twoside]{article}
\usepackage{espcrc2}
\usepackage[latin1]{inputenc}
\usepackage[dvips]{graphicx}

\title{Light Quark Masses with $N_f=2$ Wilson Fermions\thanks{Talk
    presented by N. Eicker.}}

\author{N.~Eicker\address[Wup]{Department of Physics, University of
    Wuppertal, D-42097 Wuppertal, Germany},
  Th.~Lippert\addressmark[Wup], B.~Orth\addressmark[Wup] and
  K.~Schilling\addressmark[Wup]}

\begin{document}

\begin{abstract}
  We present new data on the mass of the light and strange quarks
  from SESAM/T$\chi$L. The results were obtained on lattice-volumes of
  $16^3\times 32$ and $24^3\times 40$ points, with the possibility to
  investigate finite-size effects. Since the SESAM/T$\chi$L ensembles
  at $\beta=5.6$ have been complemented by configurations with
  $\beta=5.5$, moreover, we are now able to attempt the continuum
  extrapolation (CE) of the quark masses with standard Wilson fermions.  
  \vspace{1pc}
\end{abstract}

\maketitle

\section{Introduction}
The precise determination of light quark masses, $\overline{m}$, from
full QCD simulations is of considerable interest since the lattice
provides the only known access to their absolute scale.

We have pointed out some time ago~\cite{sesam_mass} that vacuum
polarization effects have sizeable impact on extracting $\overline{m}$
from the empirical $m_{\pi}/m_{\rho}$ ratio.  This became manifest in
a substantial ambiguity in the determination of $\overline{m}$ from
the vector Ward identities.  Indeed, there is a freedom of chiral
extrapolation ($\chi$E) on the lattice: at $\beta = 5.6.$ {\it e.g.},
the extrapolation along the line of equal valence and sea quark
hopping parameters, $\kappa^{val}= \kappa^{sea}$ (direct
extrapolation), yields a value for $\overline{m}= m_{u/d}$ which is
roughly a factor of two below the result from a semi-quenched,
two-step extrapolation procedure. In the latter one first determines
the values of $\kappa_l^{val} = \kappa_{l}^{val}(\kappa^{sea})$ for
each ensemble with fixed value of $\kappa^{sea}$ (similar to quenched
simulations).  For each $\kappa_{l}^{val}(\kappa^{sea})$
$\overline{m}^{\overline{\rm MS}}_{val}(\kappa^{sea})$ is computed
and, subsequently, this function is extrapolated to the physical point
$\kappa^{sea} = \kappa_l^{sea}$, at this given value of $\beta$.

Furthermore it has been shown by CP-PACS that the ambiguity seen at
finite $a$ goes away after CE~\cite{CPPACS}.  CP-PACS has used an
improved action that allowed them to go to rather coarse lattices,
providing them with a long lever arm for the various CE. In the
present context, their lever arm turned out to be of little value,
however, since the extrapolation with a linear CE led them to a very
poor $\chi^2$ with such an ansatz, invalidating the small error on
their value of $\overline{m}$.

Given this situation we are motivated to resume the question with the
standard Wilson fermionic action and attempt a scaling analysis, by
complementing previous SESAM/T$\chi$L data with another set of vacuum
configurations at $\beta = 5.5$, as specified in table~\ref{tab:params}.

\begin{table*}[th]
  \begin{center}
    \begin{tabular}{|c|c|c|c|c||l|l|l|l|}
      \hline
      $\beta$ & $N_s^3\times N_t$ & $\kappa$ & $N_{traj}$ & $N_{conf}$ &
      $m_{PS} /m_V$ & $(am_{PS} N_s)^{-1}$ & $r_0a^{-1}$ & $a$[GeV] \\
      \hline\hline
      5.5 & $16^3\times 32$
      &  0.1580 & 4000 & 119 & 0.854(2)  & 0.113(1) & 4.027(24) & 1.587(9)\\
      && 0.1590 & 4500 & 140 & 0.803(9)  & 0.142(1) & 4.386(26) & 1.728(10)\\
      & $a^l_{r_0}=0.092$ fm
      &  0.1596 & 5500 & 199 & 0.753(6)  & 0.170(2) & 4.675(34) & 1.842(13)\\
      && 0.1600 & 5500 & 200 & 0.675(10) & 0.204(3) & 4.889(30) & 1.926(12)\\
     \hline\hline
      5.6 & $16^3\times 32$ 
      &  0.1560 & 5700 & 198 & 0.834(3) & 0.140(1) & 5.104(29) & 2.011(11)\\
      && 0.1565 & 5900 & 208 & 0.812(9) & 0.156(2) & 5.283(52) & 2.08(2)\\
      &
      &  0.1570 & 6000 & 201 & 0.762(7) & 0.181(2) & 5.475(72) & 2.16(3)\\
      && 0.1575 & 6500 & 206 & 0.684(8) & 0.223(3) & 5.959(77) & 2.35(3)\\
      \cline{2-9}
      & $24^3\times 40$ 
      &  0.1575 & 5100 & 185 & 0.701(6)  & 0.151(2) & 5.892(27) & 2.321(11)\\
      & $a^l_{r_0}=0.076$ fm
      & 0.1580 & 4500 & 158 & 0.566(14) & 0.209(4) & 6.230(60) & 2.45(2)\\
      \hline
    \end{tabular}
    \caption{Simulation parameters. $a$-values in 2nd column hold at
      $\kappa_l$. $(am_{PS} N_s)^{-1}$ is discussed in \cite{Orth}. $r_0$
      for $\beta=5.6$ from \cite{Bolder}, for $\beta=5.5$ using the
      same approach.}
    \label{tab:params}
  \end{center}
  \vspace{-.8cm}
\end{table*}

\section{The light quark mass $\overline{m}$}

The determination of light quark masses can be sketched by the
following steps:\\
(a) Fix $\kappa_c$ from the $\chi$E of $am_{PS}^2 \to 0$;\\
(b) Get $\kappa_l$ from $\frac{m_{PS}}{m_V} \to
\frac{m_\pi}{m_\rho}$ using the $\chi$E of $am_V$;\\
(c) Determine $a^{-1}$ via $r_0 \simeq 0.5\mbox{ fm}$;\\
(d) Compute the quark mass in the lattice-scheme:
\begin{equation}
  \overline{m}^{lat}(a)=a^{-1}
  \frac{1}{2}\left(\frac{1}{\kappa_l}-\frac{1}{\kappa_c}\right);
\end{equation}
(e) Convert into $\overline{\rm MS}$-scheme using
\begin{equation}
  \label{eq:renorm}
  \overline{m}^{\overline{\rm MS}}(\mu=1/a)=Z_m\overline{m}^{lat}(a);
\end{equation}
(f) Rescale the quark mass to $\mu=2\mbox{ GeV}$;\\
(g) Extrapolate $a\to 0$.

\begin{figure}[ht]
  \includegraphics[width=.46\textwidth,
  height=.4\textwidth]{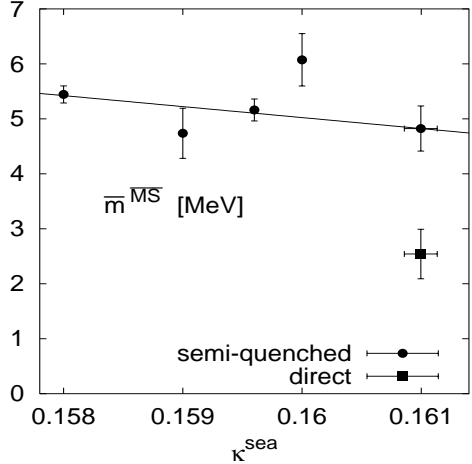}
  \vspace{-1cm}
  \caption{Light quark mass at $\beta=5.5$ in the $\overline{\rm
      MS}$-scheme at $\mu=2\mbox{ GeV}$. Square-symbol from direct,
    circles from semi-quenched extrapolation.}
  \label{fig:mlight_b5.5}
  \vspace{-0.5cm}
\end{figure}

$Z_M$ within this work was determined using a tadpole improved
perturbative procedure \cite{renorm}.

In the case of the above-quoted semi-quenched $\chi$E
procedure the results for $\overline{m}^{\overline{\rm MS}}(2\mbox{
  GeV})$ have to be extrapolated to the physical mass of the sea
quarks before sending $a\to 0$.

In order to achieve appropriate statistical errors throughout this
analysis one has to take into account the correlation between the
observables. These correlations occur especially within the
semi-quenched analysis, since all observables inside a $\chi$E are
determined on the same set of configurations.  Therefore all errors
are determined using the jackknife procedure within all steps (a) to
(g) sketched above.

\begin{figure}[ht]
  \includegraphics[width=.46\textwidth,
  height=.4\textwidth]{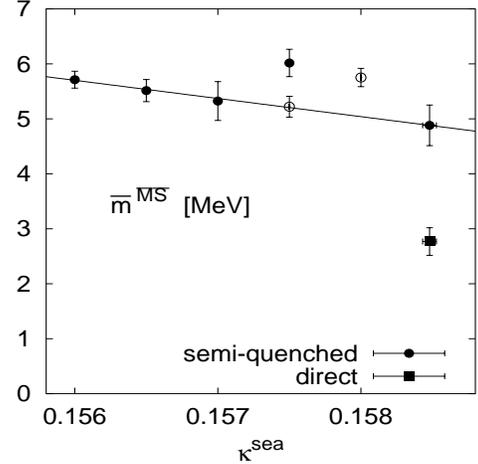}
  \vspace{-1cm}
  \caption{Same as in Fig.~\ref{fig:mlight_b5.5} but for
  $\beta=5.6$. Data with open symbols from larger lattices.}
  \label{fig:mlight_b5.6}
  \vspace{-0.5cm}
\end{figure}

Fig.~\ref{fig:mlight_b5.6} shows the results for $\overline{m}$ in the
$\overline{\rm MS}$-scheme for $\beta=5.6$ versus $\kappa^{sea}$.
Obviously the semi-quenched values for $\kappa=0.1575$ on the small
lattice and for $\kappa=0.158$ on the larger lattice deviate from the
general behavior of the data. A more detailed investigation shows
that this is due to finite-size effects. Hence these two points are
neglected in the extrapolation to the physical sea quark mass.
Furthermore we reconfirm the above mentioned discrepancy between the
results for the semi-quenched and the direct extrapolations.

Fig.~\ref{fig:mlight_b5.5} displays the corresponding results for
$\beta=5.5$. Here the value for $\kappa=0.160$ clearly shows
finite-size effects and is hence discarded.

\begin{figure}[htb]
  \includegraphics[width=.46\textwidth]{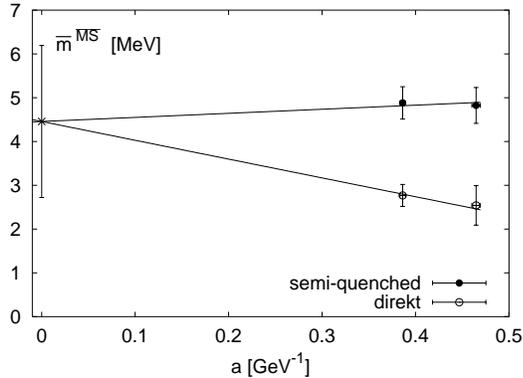}
  \vspace{-1cm}
  \caption{Extrapolating $\overline{m}$ in $a \to 0$.}
  \label{fig:mlight_vs_a}
  \vspace{-0.5cm}
\end{figure}

Fig.~\ref{fig:mlight_vs_a} presents the CE $a\to
0$, with $a$ as obtained from $r_0$ at $\kappa_l$. The continuum value
for the light quark mass is determined using a constrained fit to the
semi-quenched and direct values at two lattice constants each. The fit
is also shown in the figure.

As a final result for the light quark mass we get
\begin{equation}
  \label{eq:mlight}
  \overline{m}^{\overline{\rm MS}}(2\mbox{ GeV}) = 4.5(1.7)\mbox{ MeV}.
\end{equation}

Within the bigger errors of our analysis, this number is in good
agreement with the result quoted by CP-PACS \cite{CPPACS}.

\section{The strange quark mass $m_s$}

In contrast to the light quark mass there is no reasonable way to
carry out a direct $\chi$E for the mass of the strange quark. This is
due to the fact that directly extrapolated results would describe
mesons with strange valence quarks within a sea of strange sea quarks.
Indeed, nature is given by a sea of light sea quarks and therefore
only the semi-quenched $\chi$E makes sense.

On the other hand there are several ways to determine $\kappa_s$, the
hopping-parameter corresponding to the mass of the strange quark. In
this analysis we investigated three different inputs:
$\frac{m_\phi}{m_\rho}$, $\frac{m_K}{m_\pi}$ and
$\frac{m_{K^*}}{m_\rho}$.

\begin{figure}[t]
  \includegraphics[width=.46\textwidth]{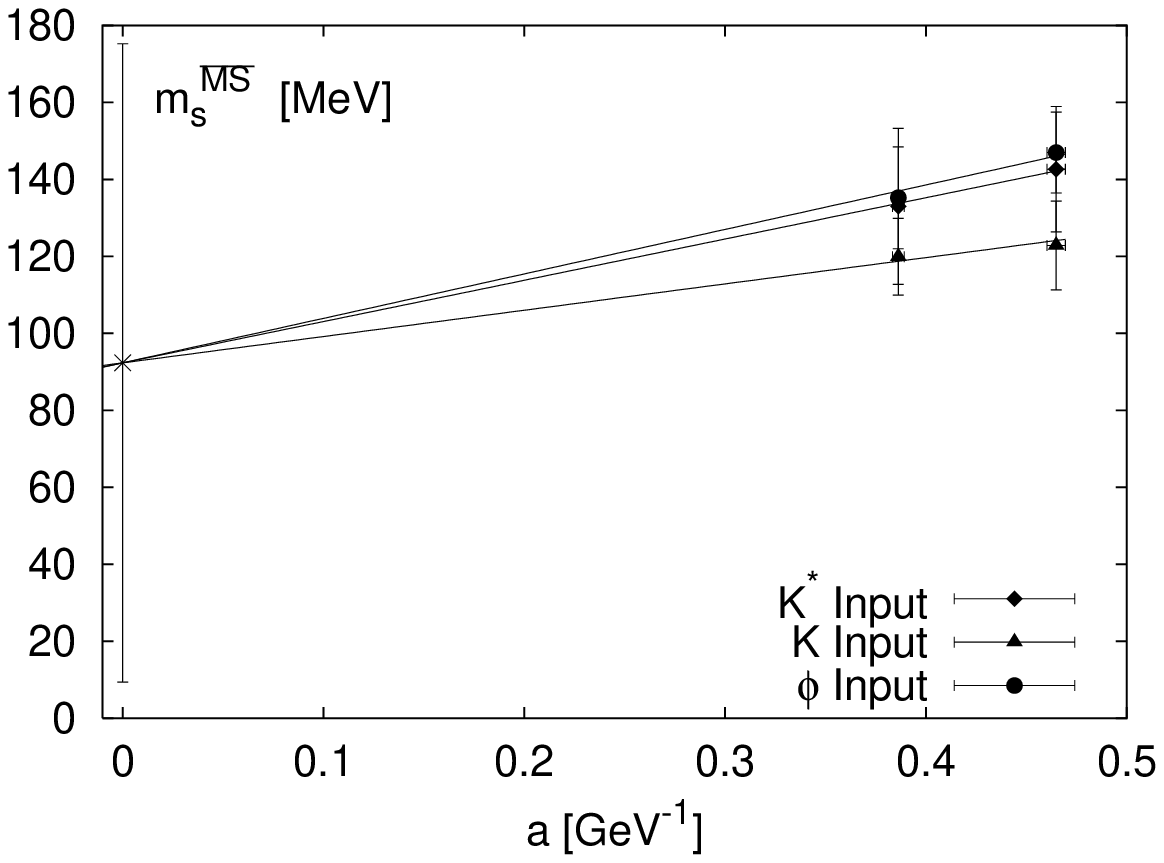}
  \vspace{-1cm}
  \caption{Extrapolating $m_s$ in $a \to 0$.}
  \label{fig:mstrange_vs_a}
  \vspace{-0.5cm}
\end{figure}

From a semi-quenched analysis of $m_s$, one obtains the results
displayed in Fig.~\ref{fig:mstrange_vs_a}.  The data clearly show
deviations for the different definitions of the quark mass which are
expected to disappear in the continuum. Hence again a constrained fit
was done for the CE.

The result of this fit for the mass of the strange quark is
\begin{equation}
  \label{eq:mstrange}
  m_s^{\overline{\rm MS}}(2\mbox{ GeV}) = 92(83)\mbox{ MeV}.
\end{equation}

Obviously the strange quark mass needs all the more
a larger range of $a$-values for a safe CE.

\section{Conclusions}
We conclude that ({\em i}) statistical errors of 5 to 10 \% can be
achieved before CE, ({\em ii}) with additional simulations at smaller
$\beta$, a reliable $\overline{m}$-determination with 10 \% accuracy
is within reach.

\section*{Acknowledgments} 
We thank G. Bali for computing the $r_0$ data, with support of the EC
under contract HPRN-CT-2000-00145 Hadrons/Lattice QCD\@.  N.E. thanks
DESY for partly supporting his PhD project. B.O. was supported by the
DFG Graduiertenkolleg ``Feldtheoretische Methoden in der
Elemen\-tar\-teil\-chen\-theorie und Statistischen Physik''.


\begin{thebibliography}{9}
\bibitem{sesam_mass} N.~Eicker, {\it et al.}, Phys.~Lett. {\bf B407}
  (1997) 290.
\bibitem{CPPACS} A.~Ali Khan {\it  et al.}, {\tt hep-lat/0105015}
  (2001).
\bibitem{Orth} B.~Orth, {\it et al.}, these proceedings.
\bibitem{Bolder} B.~Bolder {\it et al.}, Phys.~Rev. {\bf D62} (2000)
  054503.
\bibitem{renorm} P.~Lepage {\it et al.}, Phys.~Rev. {\bf D 48} (1993)
  2250.
\end{thebibliography}
\end{document}